\documentclass[11pt,twoside,a4paper]{article}  
\usepackage{graphicx,color}
\usepackage{times}
\usepackage{xspace}
\usepackage{booktabs}
\usepackage{cite}
\usepackage{epsfig}
\usepackage{feynmp}
\usepackage{a4wide}
\usepackage[small, nooneline, center]{subfigure}
\usepackage{amsmath,amssymb,bm}
\usepackage{longtable}

\usepackage[colorlinks=true, pdftex, pdftitle={Energy Scaling of Minimum-Bias Tunes}, pdfauthor={Holger Schulz, Peter Zeiler Skands}, pdfsubject={Monte Carlo generators}, pdfkeywords={Monte Carlo, generators, QCD, Pythia, Minimum Bias, HEP}]{hyperref}

\newcommand{\mrm}[1]{\mathrm{#1}}

\newcommand{\ttt}[1]{\texttt{#1}}

\newcommand{\TeV}{\,\mbox{Te\kern-0.2exV}}
\newcommand{\GeV}{\,\mbox{Ge\kern-0.2exV}\xspace}
\newcommand{\MeV}{\,\mbox{Me\kern-0.2exV}}
\newcommand{\keV}{\,\mbox{ke\kern-0.2exV}}
\newcommand{\eV}{\,\mbox{e\kern-0.2exV}}

\def\sqrts{\ensuremath{\sqrt{s}}\xspace}
\def\pt{\ensuremath{p_\perp}\xspace}
\definecolor{todoboxcolor}{rgb}{1.0,0.6,0.6}

\definecolor{todoneboxcolor}{rgb}{1.0,0.9,0.9}

\newcommand{\chisq}{\ensuremath{\chi^2}\xspace}
\newcommand{\p}{\ensuremath{\vec{p}}\xspace}

\newcommand{\wb}{\ensuremath{w_{b}}\xspace}

\def\parp#1{\ensuremath{\mathtt{PARP(#1)}}\xspace}

\newcommand{\EqRef}[1]{equation~\eqref{#1}\xspace}
\newcommand{\SecRef}[1]{section~\ref{#1}\xspace}
\newcommand{\TabRef}[1]{Table~\ref{#1}\xspace}
\newcommand{\FigRef}[1]{Figure~\ref{#1}\xspace}

\def\professor{\textsc{Professor}\xspace}
\def\rivet{Rivet\xspace}

\begin{document}
\begin{minipage}{\textwidth}
\flushright
CERN-TH-2011-036\\
MCNET-11-07 \\
HU-EP-11/13
\end{minipage}
\vskip5mm
\begin{center}
{\Large Energy Scaling of Minimum-Bias Tunes}
\end{center}
\vskip5mm
\begin{center}
{\large H.\ Schulz$^1$, P.\ Z.\ Skands$^2$}
\end{center}
\begin{center}
\parbox{0.9\textwidth}{
$^1$: Institut f\"ur Physik, Humboldt-Universit\"at zu Berlin, Germany\\
$^2$: Theoretical Physics, CERN, CH-1211 Geneva 23, Switzerland
}
\end{center}
%\maketitle
\vskip5mm
\begin{center}
\parbox{0.85\textwidth}{
\begin{center}
\textbf{Abstract}
\end{center}\small
We propose that the flexibility offered by modern event-generator tuning tools 
allows for more than just obtaining ``best fits'' to a collection of
data. In particular, we argue that the universality of the underlying
physics model can be tested by performing several, mutually
independent, optimizations of the generator parameters in different
physical regions. For regions in which these optimizations return
similar and self-consistent parameter values, the model can be
considered universal. Deviations from this behavior can be associated
with a breakdown of the modeling, with the nature of the deviations
giving clues as to the nature of the breakdown. We apply this
procedure to study the energy scaling of a class of minimum-bias
models based on multiple parton interactions (MPI) and
$p_\perp$-ordered showers,  
implemented in the \textsc{Pythia} 6.4 generator. We find that a
parameter controlling the strength of color reconnections in the final
state is the most important source of non-universality in this model.  
}
\end{center}

\section{Introduction}
The main virtue of general-purpose Monte Carlo event generators
(sometimes called ``shower'' Monte Carlos, although they are normally
relied on for many other physics aspects as well) is their ability to
provide a complete and fully differential picture of collider final
states, down to the level of individual particles. This  
allows them to be used as detailed --- albeit approximate --- 
theoretical references for measurements performed at accelerators like the LHC, 
against which models of both known and `new' physics can be tested. 
The achievable accuracy depends both on the
inclusiveness of the chosen observable (with more
inclusive  observables
generally being more precisely predicted) and on the  
sophistication of the simulation itself. An important driver for the
latter is obviously the development of improved theoretical models,
e.g., by including matching to higher-order matrix elements, more
accurate resummations, or better non-perturbative models; but it  
also depends crucially on the available constraints on the remaining
free parameters of the model. Using existing data to constrain these is
referred to as generator tuning.  

The recent minimum-bias measurements from the LHC experiments, in
particular those at centre-of-mass energy $\sqrt{s} = 7$ TeV, 
have highlighted the question to what extent the energy
scaling of total cross sections and differential distributions 
are consistent with model-based extrapolations from lower energies, or
whether they exhibit any non-trivial departures from such
predictions. 
Most of the LHC collaborations have already gone some way
towards studying this question, by including comparisons of specific
models and tunes to the data in their publications. In the short term,
such comparisons are useful both as immediate  tests of commonly used
models, and to illustrate the current amount of theoretical
uncertainty surrounding a particular distribution.  They also 
provide a set of well-defined theoretical reference 
curves for future studies. However, the conclusions that can be drawn
from comparisons of individual tunes of specific models on single
distributions are necessarily limited. In order to obtain more general
conclusions, a more coherent and over-arching look at both the data
and the models is needed. 

In this study, we shall make use of the \professor tuning tool
\cite{Buckley:2009bj} to provide such a look. Specifically, rather than
performing one global fit to all the data, as is usually done, we
 instead use \professor to perform several independent
optimizations of the model parameters for a range of different collider
energies. At each energy, we use the same set of minimum-bias
observables, modulo the limitations imposed by
different detector acceptances, trigger
conditions, and correction procedures. 
We thereby seek to obtain a data-driven map of the preferred energy dependence 
of each of the tuned parameters. This can then 
be compared to the functional dependence assumed
by the underlying model, thereby furnishing  not just a ``best fit'' of the model
parameters, but also a consistency check on the universality of the
underlying physics model itself.

We emphasize that this sort of consistency check is not limited to
energy scaling alone. In all generality, a consistency check 
on the underlying physics model can be obtained by performing 
independent optimizations in any two (or more) ``physics windows'' 
that the modeling provides or assumes relations between.
Moreover, with the recent advent of  
automated tuning tools, it is now becoming possible to explore many 
such independent optimizations with only modest investments 
of computing and manpower. 
In regions in which consistent parameter sets are obtained, with 
predictions that are acceptably close to the data,
the model can be considered as interpolating well, i.e., it is universal. 
If not, a breakdown in the ability of the model ability to span different
physical regimes has been identified, and can be addressed.

For simplicity, we concentrate on one particular model here, the
`new' interleaved multiple-interactions model
\cite{Sjostrand:2004pf,Sjostrand:2004ef}  
implemented in the \textsc{Pythia 6} event generator \cite{Sjostrand:2006za}. 
All parameters not explicitly subjected to optimization in this study
are those of the `Perugia 0' tune
\cite{Bartalini:2010su,Skands:2010ak}. We note that this model has
significant similarities with, but is not identical to, the one
implemented in \textsc{Pythia} 8 \cite{Sjostrand:2007gs}. 

In section \ref{sec:setup}, we give brief overviews of the theory model, the
\professor/\rivet tuning framework, and the data sets we have used to do
the tuning. Section \ref{sec:energyscaling} contains our main
study of the energy scaling of three main model parameters at
currently existing colliders. 
We round off with conclusions and outlook in section \ref{sec:conc}.

\section{Setup \label{sec:setup}}
In this section, we briefly describe the theoretical models we take as
starting points, emphasizing in particular
the parameters relevant to this study  (section \ref{sec:model}), the tuning
framework we will be 
using (section \ref{sec:professor}), and the data
sets that have been included (section \ref{sec:datasets}).

\subsection{Theoretical Model \label{sec:model}}

As mentioned above, 
we consider the interleaved model of $p_\perp$-ordered showers and
multiple parton-parton interactions of \cite{Sjostrand:2004pf,Sjostrand:2004ef}, 
as implemented in \textsc{Pythia} 6.4.23 \cite{Sjostrand:2006za}, 
specifically the `Perugia 0' tune of that model
\cite{Skands:2010ak} unless otherwise specified. 

In this class of models, pioneered by \cite{Sjostrand:1987su}, a
unified approach is taken to the modeling of all inelastic
non-diffractive events, in which dijet production and its associated
underlying event (UE) is viewed merely as the hard high-$p_\perp$ tail of
minimum-bias (MB), without any sharp modeling distinction 
between the two. The fundamental building block for the model is the dijet
cross section, computed at leading order in perturbative QCD, 
whose dominant component is simple low-$p_\perp$ Rutherford scattering
($t$-channel gluon exchange). 

For $p_\perp\to 0$, i.e., when two partons  scatter via the exchange
of a very soft gluon, this cross section
exhibits a divergence whose ultimate origin is similar to that of 
initial-state bremsstrahlung. And similarly to what is done for 
bremsstrahlung, the model recasts this divergence in terms of a
unitarized (Sudakov-suppressed) finite cross section for the hardest
scattering accompanied by a divergent \emph{number} of successively softer
multiple parton interactions (MPI), a number which is ultimately
regulated by the introduction of an infrared regularization scale.  We do
not intend to give a full account of unitarization here, but instead
refer the reader to
\cite{Sjostrand:1987su,Sjostrand:2004pf,Sjostrand:2006za} for details
on it in the context of the MPI models here discussed. We also note
that an interesting exploration of the relation between
this effective scale and perturbative low-$x$ physics was recently 
carried out
in \cite{Ryskin:2011qe}. Finally, a pedagogical
and more general discussion of underlying-event models in
general-purpose Monte Carlo generators can be found in the recent
review \cite{Buckley:2011ms}. 

In this study, we focus on three main parameters of the resulting type
of model: the infrared
regularization scale, the proton transverse mass distribution, and the 
color-reconnection strength, as follows:

\paragraph{Infrared Regularization Scale:} The fact that long-wavelength gluons
only see a coherent sum of the color charges in the hadronic
substructure --- color screening --- is assumed to ultimately regulate
the divergent number of parton-parton scatterings, similarly to how
the non-perturbative cutoff in parton showers regulates the
number of parton shower emissions. In the model we consider here,
a smooth regulator is introduced, by modyfing 
the divergent parts of the cross section (including the strong
coupling since we use the standard MC scale choice
$\alpha_s(p_\perp^2)$) as follows,  
\begin{equation}
\alpha_s(p_\perp^2) \ \frac{\mathrm{d}p_\perp^2}{p_\perp^4} \ \to \ 
\alpha_s(p_{\perp0}^2 + p_\perp^2) \ 
\frac{\mathrm{d}p_\perp^2}{(p_{\perp0}^2 + p_\perp^2)^2} ~,
\end{equation}
where $p_{\perp 0}$ physically expresses the scale at which the color screening
effect is supposed to become active. This parameter, which we call the
{\bf infrared regularization scale}, constitutes the main free 
parameter for all models of this type, with low values yielding more
soft MPI activity (in the limit that it is taken to zero, the original
unregulated behaviour would be reobtained). In the \textsc{Pythia} model,
it is assumed to have a power-law scaling with the CM energy, 
\begin{equation}
    \label{eqn:ptmin}
    p_{\perp 0}(\sqrts) = \parp{82}\cdot
    \left(\frac{\sqrts}{\parp{89}}\right)^\parp{90}~, 
\end{equation}
where \parp{82}, \parp{89}, and \parp{90} are tunable parameters. 
Roughly speaking, \parp{82} gives the value of $p_{\perp 0}$
(in GeV) at a fixed reference CM energy = \parp{89} (also in GeV), and \parp{90}
determines the scaling behaviour of $p_{\perp0}$ away from that
energy. Below, instead of assuming the form, eq.~(\ref{eqn:ptmin}), we
shall fit for  $p_{\perp0}$ independently at several different values
of $\sqrt{s}$. (Technically, we do this by fixing \parp{89} to the
energy of the relevant collider and fitting for \parp{82} which can
then be interpreted directly as $p_{\perp0}$ at that energy.) We can then
check whether the resulting points lie on a curve that is 
consistent with the functional form of eq.~(\ref{eqn:ptmin}) or not. 

\begin{figure}[t]
\vspace*{-0.1cm}\hspace*{-0.33cm}\includegraphics[scale=0.42]{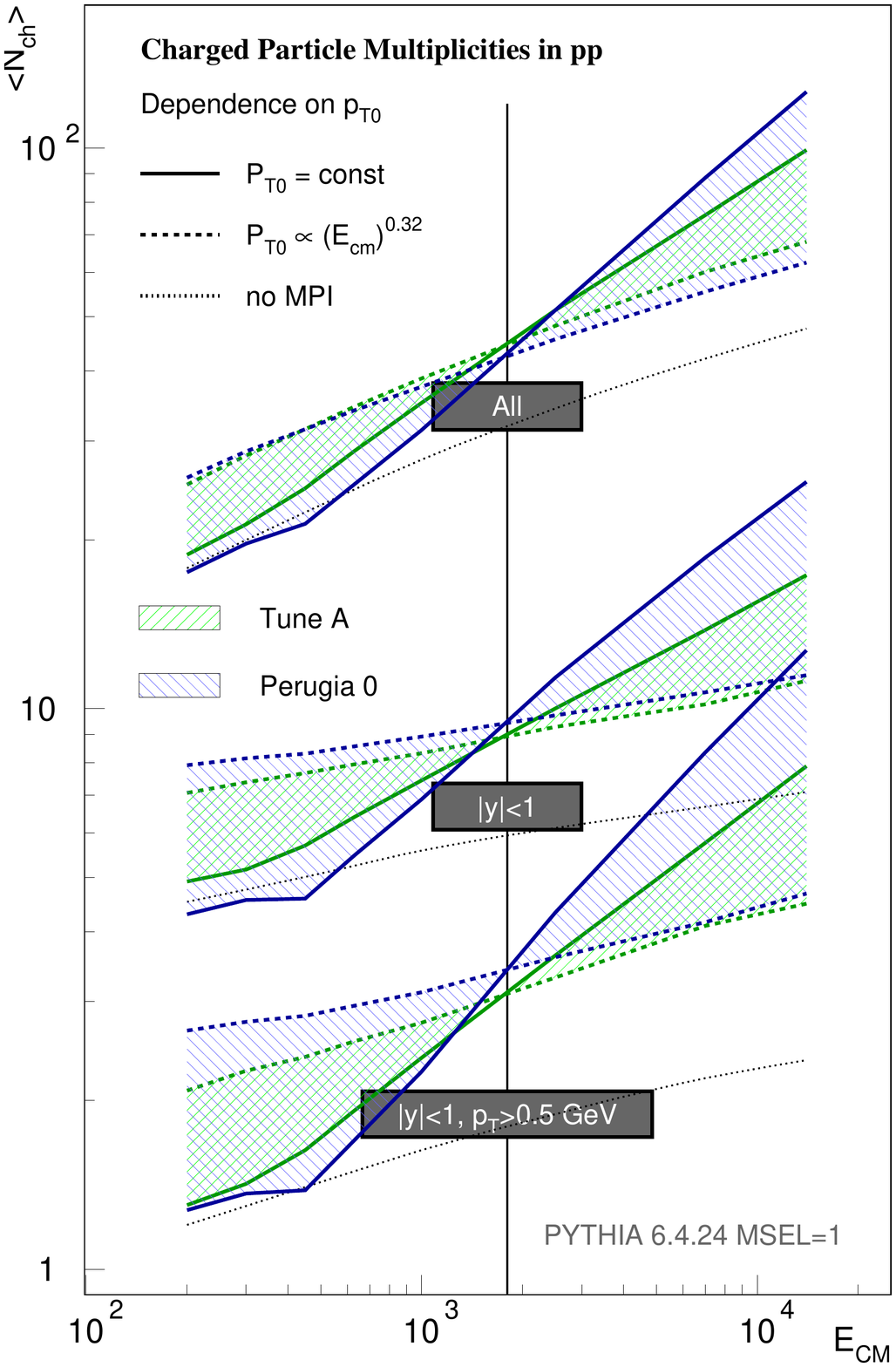}\hspace*{-1.0cm}\includegraphics[scale=0.42]{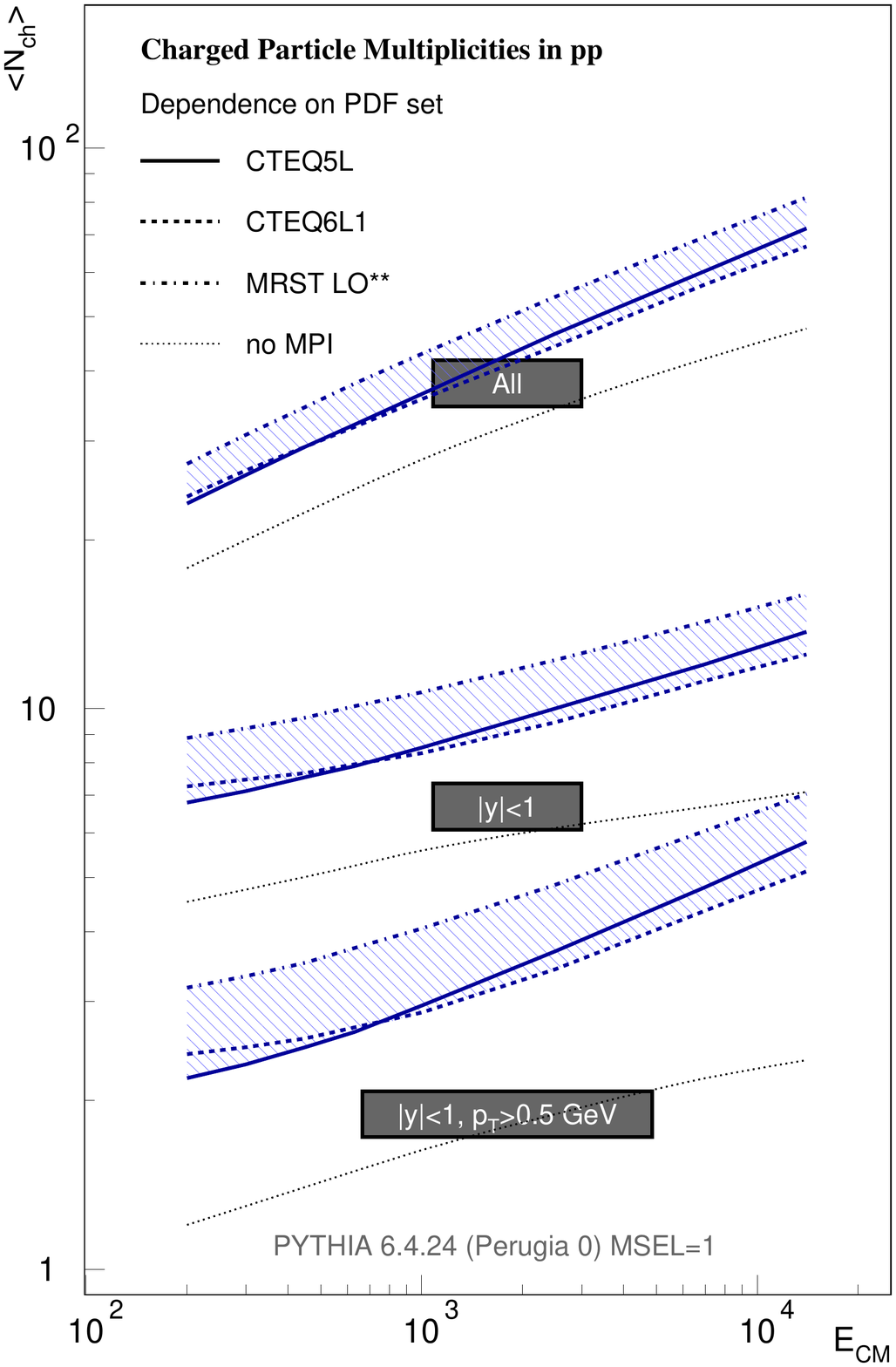}\vspace*{-0.7cm}
\caption{Energy scaling of charged-particle multiplicities in
  pp in three different phase space
  regions (top: inclusive, middle: central, bottom: central hard). 
  {\sl Left:} Dependence on the scaling of the $p_{\perp 0}$ parameter for
  two different \textsc{Pythia} models, represented by Tune A and
  Perugia 0, respectively. The solid vertical line represents the reference energy,
  1800 GeV, at which \ttt{PARP(82)} is defined for both models.
  {\sl Right:} Dependence on the PDF set, for
  the Perugia 0 model.  
  For reference, Tune A without MPI is also shown (dotted
  lines). 
\label{fig:th-pt0}} 
\end{figure}

To give the reader a more concrete idea of the dependence of the overall 
event activity on the assumed scaling form, the left-hand pane of
Fig.~\ref{fig:th-pt0} illustrates the scaling
behaviour of 
charged-particle multiplicities in non-diffractive minimum-bias
events\footnote{Specifically, the generated
events correspond to running \textsc{Pythia} in its ``minimum-bias''
mode with diffraction switched off. We permit ourselves this somewhat
unphysical definition here, since the illustration is intended for
qualitative purposes only. (For the numerical
studies later in this report, we use a full inelastic sample that
includes diffraction.)},  
for two different
assumptions of the energy scaling of the $p_{\perp 0}$ parameter which we consider
comparatively extreme: 
\begin{enumerate} \item 
Solid lines: constant $p_{\perp 0}$, i.e., \parp{90} = 0.0, resulting 
in a very fast growth of multiplicity with energy.
\item Dashed lines:  \parp{90} = 0.32, i.e., $p_{\perp 0}$ varying 
as $(\sqrts)^{0.32}$, resulting in a
multiplicity growth with energy which is comparable to the 
case without MPI, shown with dotted lines.
\end{enumerate} 
For completeness, both
of the two \textsc{Pythia} min-bias models are included here, represented
by Tune A and Perugia 0, respectively. (For reference, the default for Tune A is a
scaling power of 0.25. For Perugia 0, it is 0.26.) In both cases, three
phase space regions are shown: inclusive (top), central (middle), and
central hard (bottom), with phase space cuts as indicated in the
grey shaded boxes.
Comparing model curves with equal scaling assumptions (solid with
solid and dashed with dashed), it is evident that the two models have
somewhat different intrinsic scaling properties, 
even with the same assumption for the
scaling of $p_{\perp 0}$. Thus, the value and scaling of $p_{\perp 0}$
alone is not sufficient to fix the energy scaling completely. 

It is also worth noting that both models require \emph{at least one}
partonic scattering per hadron-hadron collision, hence it is
not possible for the average multiplicity to drop below the 
``no-MPI'' case. This causes a rather abrupt change
in the scaling behaviour at low energies for those curves that
intersect the no-MPI (dotted) one. 

On the right-hand pane of Fig.~\ref{fig:th-pt0}, we illustrate another
important dependence, on the PDF set used. Both Tune A and Perugia 0
were made using the CTEQ5L set \cite{Lai:1999wy} (solid
lines). Changing to either  
CTEQ6L1 \cite{Pumplin:2002vw} (dashed lines) or MRST
LO** \cite{Sherstnev:2007nd} (dot-dashed lines)  affects both the
value of the average 
multiplicities as well as their scaling behaviour and distribution in
phase space. In particular, the LO** set generates a
significantly larger activity than its CTEQ cousins. It is therefore
not generally possible to separate the choice of PDF set from the
$p_{\perp 0}$ choice. 

\paragraph{Transverse Mass Distribution:} A further important aspect
of the model is the  shape of the assumed  
proton matter distribution. In MPI models, 
the probability for 
additional parton-parton interactions to occur in a given collision is
proportional to the amount of matter overlap between the colliding beam
particles in that collision, which in turn depends on their impact
parameter, $b$. If the proton structure is very uniform (e.g., a
featureless pion/gluon cloud), the differences between peripheral and central
collisions will be quite small, while a strongly peaked distribution
(e.g., valence lumps / hot spots) can make the activity in
central collisions much higher than in peripheral ones. Thus,
while we may think of the infrared regularization scale above as 
determining the \emph{average} number of multiple parton interactions, 
the $b$ profile affects how much this number can \emph{deviate} 
from the mean in peripheral vs.\ central events. 
In the overlap model used for the
Perugia tunes, the overlap function (the time-integrated convolution
of two proton mass distributions, see
\cite{Sjostrand:1987su,Sjostrand:2004pf}) is cast as
\begin{equation}
{\cal O}(b) \propto \exp\left(-b^d\right) 
\end{equation}
with the power $d$ a free parameter whose range is normally taken to
be from $d=1$ (exponential, representing a very peaked structure) 
to $d=2$ (Gaussian, representing a smooth structure). Note that the
normalization of this distribution is fixed to unity. Note also that
$b$ is given in an arbitrary unit; since the only dimensionful
quantity is the total cross section, which is fixed by a 
Donnachie-Landshoff formula \cite{Donnachie:1992ny}, the $b$ shape
does not affect the total cross section at all in this type of model, 
and only the dimensionless ratio $b/\left<b\right>$ appears in the 
explicit calculations\footnote{
For completeness, we note that, 
while there is thus formally a dependence on the
overall proton-proton impact parameter $b$ in
the model, there is no actual space-time representation of the
collision, and hence no dependence on the direction of $b$ nor on the
individual parton-parton impact parameters. }. 

 The power, $d$, appears as the parameter \parp{83} in 
 \textsc{Pythia}\footnote{Strictly speaking, this form of the matter
   profile is only selected for 
   \ttt{MSTP(82)=5}. See the \textsc{Pythia} documentation on
   \ttt{MSTP(82)} for how to select other matter profiles, 
   such as the double-Gaussian one \cite{Sjostrand:2006za}.}. 
It is
 not assumed to change with energy, i.e., 
\begin{equation}
d(\sqrt{s}) = \parp{83} ~.
\end{equation}
By making separate tunes at each energy individually, we will obtain a
 data-driven test of the validity of this assumption.

\begin{figure}[t]
\vspace*{-0.1cm}\hspace*{-0.33cm}\includegraphics[scale=0.42]{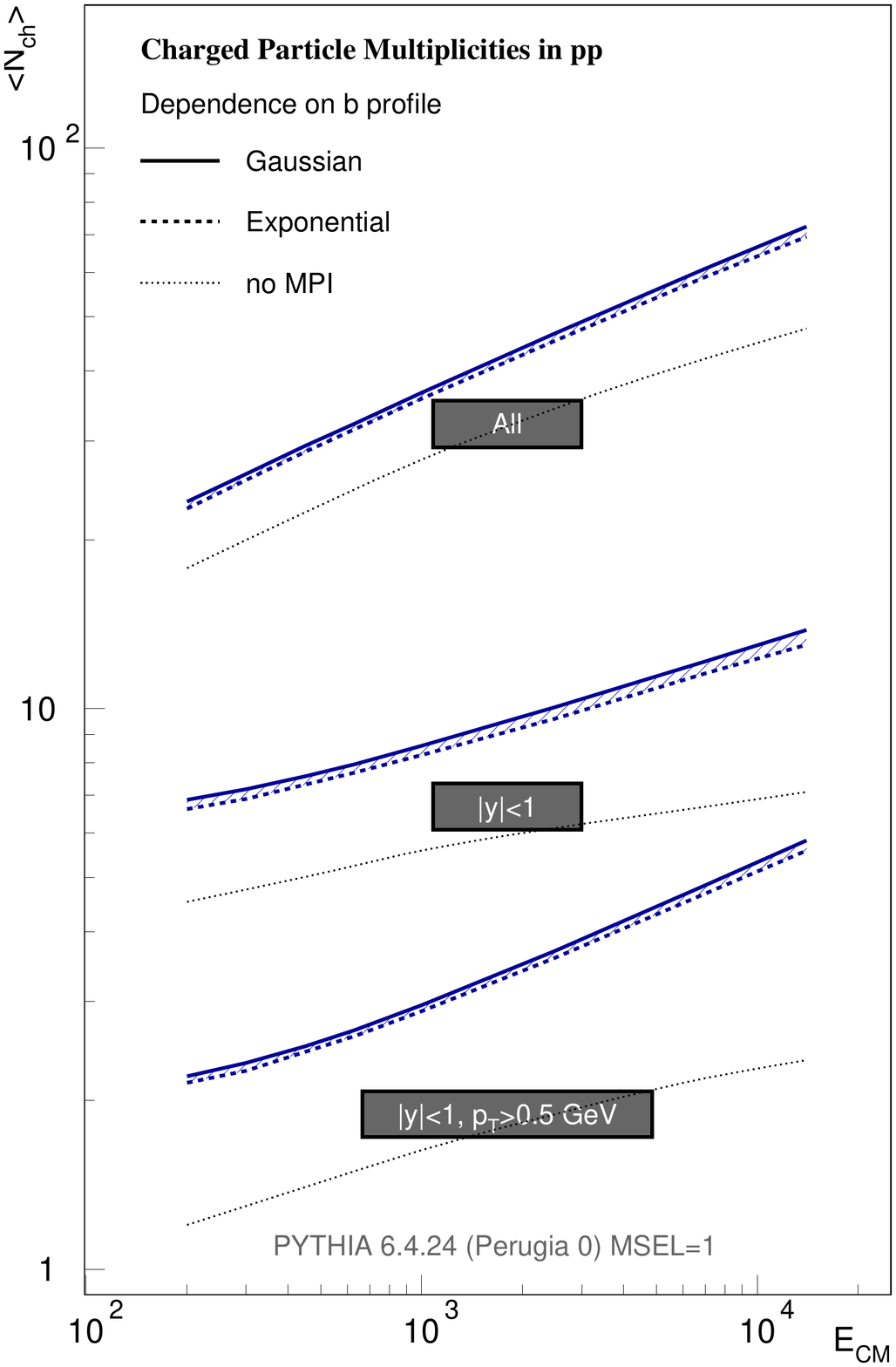}\hspace*{-1.0cm}\includegraphics[scale=0.42]{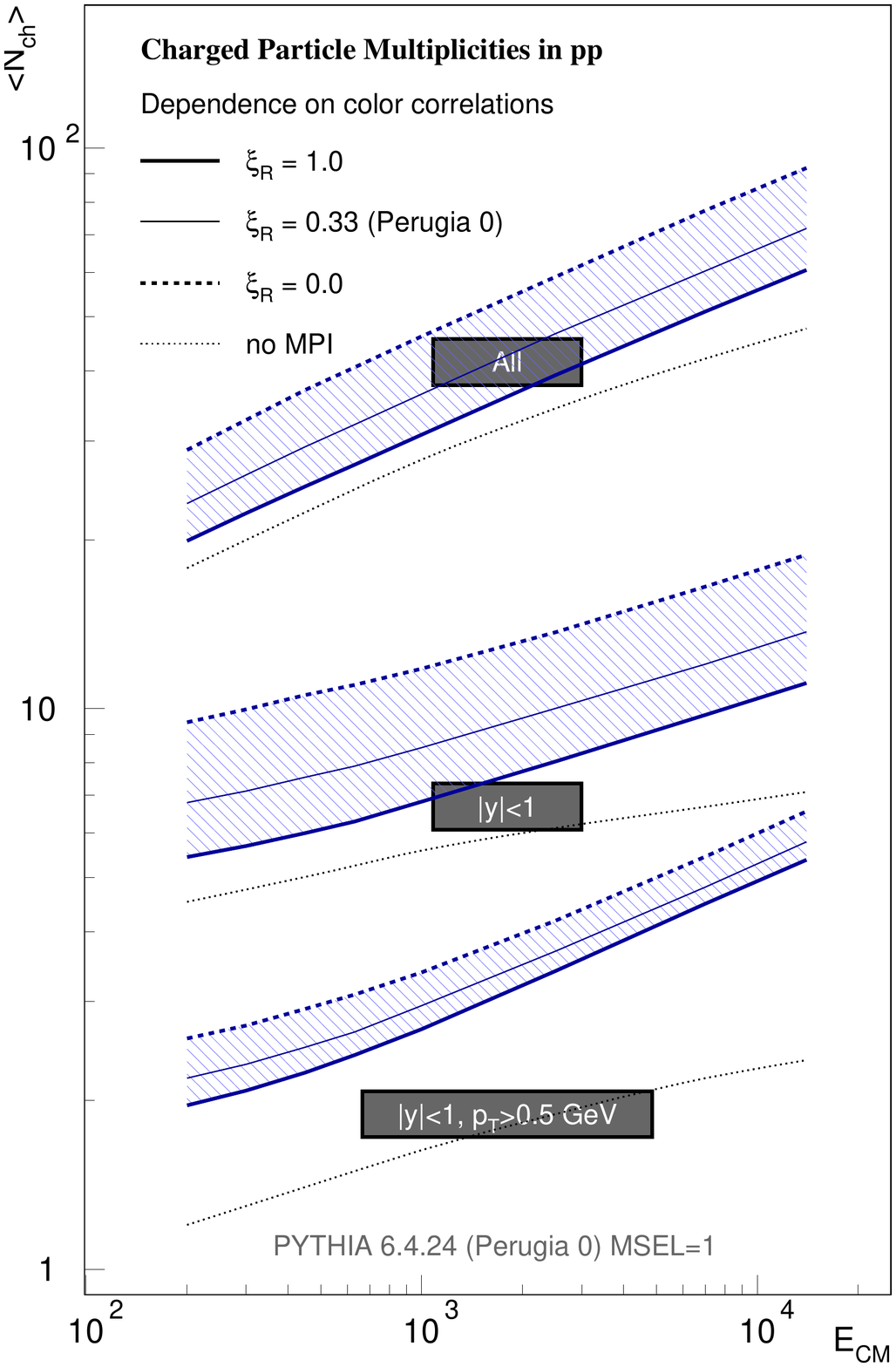}\vspace*{-0.7cm}
\caption{Energy scaling of charged-particle multiplicities in
  pp in three different phase space
  regions (top: inclusive, middle: central, bottom: central hard). 
  {\sl Left:} two different impact parameter profiles. {\sl Right:}
    three different color-reconnection strengths. 
  For reference, Tune A without MPI is also shown (dotted
  lines). For all other curves, the parameters of Perugia 0 were used,
  except for the modifications indicated on the plots.
\label{fig:th-b}} 
\end{figure}

Since all expressions are cast in terms of the
dimensionless ratio of the impact parameter relative to its average, 
the assumed shape also does not greatly affect the \emph{average} event
activity. The main consequence of different $b$ profiles thus lies in the
\emph{shape} of distributions, with a smooth matter profile generating
narrower ones than more lumpy profiles, a consequence of the latter
allowing for larger
event-to-event fluctuations. The fact that the
average multiplicity is not greatly affected by the choice of
impact parameter profile is illustrated on the left-hand pane of 
Fig.~\ref{fig:th-b},  
where the $b$ dependence of the Perugia 0 tune has been varied between a
Gaussian and an Exponential overlap distribution without substantially
altering neither the average values nor their scaling with energy. 

\paragraph{Color Reconnection Strength:} The last main aspect of the
modeling we shall be concerned with here is 
the strength of the color reconnections (CR) that are used to model the
collapse of the color wave function in the final state. The 
so-called `color annealing' models employed by the Perugia tunes
were described in detail in
\cite{Sandhoff:2005jh,Buttar:2006zd,Skands:2007zg,Skands:2010ak} and are
qualitatively similar to the Generalized-Area-Law (GAL) models
developed earlier by the Uppsala group \cite{Rathsman:1998tp}. Briefly summarized,
each MPI corresponds to one or two color exchanges between the
beams (depending on whether a quark or a gluon is exchanged,
respectively). 
In the $N_C\to\infty$ limit used to represent color topologies
in MC generators, every such exchange must be neutralized at
the hadronization stage. In \textsc{Pythia}'s case this is modeled 
by the formation of
strings, which subsequently hadronize to produce observable
particles. 

In the most naive $N_C\to\infty$ treatment, each such string would be
completely independent of the others. However, since the number of
real-world colors is finite, $N_C=3$, and since
the strings generated by MPI all traverse the same rapidity region
(between the remnants) there is some reason to suppose that the collapse of the
color wavefunction is instead more complicated and/or that
the strings after formation interact to fuse or cut each other
up. In the string picture, such effects should be driven
by a minimization of the total space-time area spanned by the strings
(the so-called area law for classical strings, as measured, e.g.,
by the $\lambda$ measure \cite{Andersson:1983jt,Andersson:1985qr}). Even without
understanding the dynamics in detail, we may therefore reasonably suspect that
the end result will be shorter string pieces, which in turn
will produce 
fewer, but more energetic, particles, i.e., a harder fragmentation
spectrum. 

At least within the $p_\perp$-ordered \textsc{Pythia} 6
modeling, some such mechanism does appear to be empirically
necessary in order to properly describe the observed increase of the
mean $p_\perp$ of 
charged tracks with track multiplicity in min-bias events 
\cite{Aaltonen:2009ne,Aad:2010rd}. 

The annealing models developed in
\cite{Sandhoff:2005jh,Buttar:2006zd,Skands:2007zg,Skands:2010ak} are all formulated
in terms of one main
parameter: the basic color-reconnection/string-interaction
strength, $\xi_R$, given by \parp{78} in the code. The larger this parameter
is, the stronger the reconnection effect, and the faster the rise
of  $\left<p_\perp\right>(N_{ch})$. However, since these models were only
intended as crude toy models, nothing has so far been said as to their
possible dependence on the energies of the colliding beams. The only
scaling built into the models is thus a rough scaling with the number
of MPI in an event, or in the most detailed variant (so far used only for the
Perugia 2010 and Perugia K tunes \cite{Skands:2010ak}) the number of
overlapping string pieces in each rapidity region. The fundamental
reconnection probability is assumed constant, i.e., 
\begin{equation}
\xi_R(\sqrt{s}) = \parp{78}~.
\end{equation}
Again, by making separate tunes at each energy individually, we will obtain a
 data-driven test of the validity of this assumption.

The consequence of varying \parp{78} from zero to one is illustrated
in the right-hand pane of Fig.~\ref{fig:th-b}. We observe that the
average multiplicity at each energy can be modified by up to a factor
of 2 by this particular colour-reconnection model, 
but note also that the relative scaling \emph{between} energies stays
virtually independent of $\xi_R$. Nonetheless, since any model with,
for instance, an energy-dependent $\xi_R$ would interpolate between our
curves --- leading to a different effective energy dependence --- 
we must still conclude that the energy scaling of other models
could be qualitatively different from this.

Finally, we note 
that the CR model employed by the Perugia 0 tune actually depends on one
more parameter, \parp{77}, which acts to suppress reconnections among
high-$p_\perp$ string pieces. Since this parameter is tightly
correlated with \parp{78} and since it only affects details of the
high-$p_\perp$ tail of the $p_\perp$ distribution, we have kept it
fixed to its Perugia-0 value in this study. However, we did check that
  allowing its value to change or even fixing it to zero did not 
qualitatively alter any of our conclusions.

\paragraph{Remarks on Diffraction:}
Finally, we should emphasize a point that is especially relevant for 
the modeling of low-multiplicity minimum-bias collisions. As
mentioned above, the model we have outlined here attempts primarily 
to describe
inelastic, non-diffractive events. 
It would therefore have to be complemented by a separate
modeling of the diffractive parts of the cross section, to the extent that the
measurements are sensitive to these, as, e.g., in low-multiplicity
minimum-bias events. In this study, we use the diffractive modeling 
provided by the \textsc{Pythia} 6 generator itself. We note that this
modeling does not include diffractive jet production and is presumably
not reliable for high-$p_\perp$ and/or high-mass diffractive particle
production. We therefore investigate the energy scaling of the
resulting combined min-bias model for two subsamples of events --- one that
includes the low-multiplicity region, and one that excludes it, as
will be discussed in the description of the data sets below.

\subsection{Tuning Framework \label{sec:professor}}
The \professor tuning framework~\cite{Buckley:2009bj} relies on the construction of a
fast analytic model of the generator by bin-wise parameterizations of the
generator's response to shifts in parameter-space. These
parameterizations are performed within a hypercube of user-specified
volume inside the generator parameter space. (It is up to the user to
make sure that the boundaries of this hypercube correspond to
meaningful generator settings.) The comparison to experimental
analyzes is performed via the \rivet \cite{Waugh:2006ip,rivet} analysis tool, 
which in turn relies on the \textsc{HepMC} \cite{Dobbs:2001ck} event
record format and on the \textsc{HepData} repository \cite{Buckley:2010jn}.

An important point is to choose analyzes in \rivet that contain
observables that are sensitive to the tuning-parameters. The \professor system
contains tools that help to identify those observables that are not sensitive
to the parameters in question (\ttt{prof-sensitivities}) and to confirm that
the sampling-hypercube is chosen in such a way that the Monte-Carlo-generator
runs enclose the experimental data (\ttt{prof-envelopes}). 

The actual tuning stage consists of a numerical minimization of a
goodness-of-fit (GOF) measure
constructed from the parameterizations $f^{(b)}(\p)$, experimental data $\mathcal{R}_b$
and most important a weight \wb for each bin of a set of observables to tune to:
\begin{align}
  \chisq(\p) = 
  \sum_{\mathcal{O}} \sum_{b \, \in \, \mathcal{O}} 
  \wb\cdot\frac{ (f^{(b)}(\p) - \mathcal{R}_b)^2 }{ \Delta^2_b },
  \label{eq:chi2}
\end{align}%
The bin-weights \wb can be seen as the main user-input to the tuning-stage; they
help to emphasize or exclude certain regions of an observable in the
numerical fitting procedure.  The return value of the fit will be a point \p that
minimizes the GOF given in \EqRef{eq:chi2}. This parameter set may then 
be subjected to explicit validation in terms of comparison of the
observable as predicted from the parameterizations at \p 
with the experimental data. Finding a set of weights that leads to a
satisfying description of observables usually requires some iteration.

For each collider energy, we perform a ``local'' tune that includes
only data from that particular energy. For these, we manually set
\parp{89} equal to the given energy and let \textsc{Professor} optimize for
\parp{82}, which can then be interpreted directly as a $p_{\perp 0}$
value at that energy.  We used the Perugia 0 parameter settings as center of our
parameter sampling hypercube, and used the Perugia 0 energy scaling to
define the sampling range for \parp{82} at each energy. 
An overview of the parameter sampling-ranges is given in
\TabRef{tab:ranges}. 
\begin{table}[t]
\centering
\scalebox{0.9}{
\begin{tabular}{l*{5}{c}|c}
\toprule
\sqrts & 630 \GeV & 900 \GeV & 1.8 \TeV & 1.96 \TeV & 7 \TeV & Global fit\\
\midrule
\parp{82} & 1.066 \ldots 1.979 & 1.169 \ldots 2.171 & 1.4 \ldots 2.6 & 1.431 \ldots 2.658 & 1.993 \ldots 3.702 & 1.0 \ldots 3.0\\
\midrule
\parp{78} & \multicolumn{6}{c}{0.0 \ldots 0.7}  \\
\parp{83} & \multicolumn{6}{c}{1.0 \ldots 2.0}  \\
\midrule
\parp{90} & &&&& & 0.16 \ldots 0.34 \\
\bottomrule
\end{tabular}}
\caption{Parameter sampling-ranges used for the MC generator runs at each energy.}
\label{tab:ranges}
\end{table}

We also perform a ``global'' tune, using data from all energies
simultaneously, with an approximate relative weighting that attempts
to take into account that the amount of data --- and hence the
statistical power --- at each energy varies. The number of events
contained in each data sample we used is listed in
\TabRef{tab:nevt}. (These data sets will be discussed in more detail in
\SecRef{sec:datasets}.) In order to take into account the possibility 
that several measurements of the same observable may have been 
performed at closely spaced energies (e.g., Tevatron Run I and II), 
we define an effective total number of events for each observable, for each
collider energy, as follows:
\begin{equation}
N_i^{\mrm{eff}} = \sum_j N_j \, e^{-r_{ij}^2/(2\sigma^2_E)}~,
~\label{eq:Neff}
\end{equation}
where $j$ runs over all included measurements of the given observable
at all energies, $r_{ij}=\log_2(E_i/E_j)$ provides a
logarithmic measure of the distance between two energies, 
and we have chosen an ``energy resolution parameter'' of $\sigma_E =
1/3$ so that energies spaced a factor of 2 or more apart correspond to
being $3\,\sigma$ away from each other and will therefore effectively 
contribute  independently, while measurements closer than a
factor $2^{1/3}\sim1.25$ in energy will blend into each 
other, being resolved by less than $1\,\sigma$.
(We note that this parameter could be
varied to help estimate the uncertainty on the tuning, but the
question of more rigorous uncertainties is not a simple one and 
reaches beyond the scope we aim to address here.) 
The effective event numbers computed in this way are 
given, for each observable, in the two rightmost columns of
\TabRef{tab:nevt}, with the figures below illustrating the effective 
contributions of each sample to the total at each energy. 

The most naive weight normalization would be to simply let all the
samples enter with unit weights. This would unavoidably 
bias the fit towards the energies at which most statistics has been
collected. Using $N_i^\mrm{eff}$, we may instead attempt to normalize the
statistical power in such a way as to force each energy to enter with
approximately equal weight, regardless of the amount of statistics
collected. This could be achieved by weighting each
sample by
\begin{equation}
w_i^{\mrm{eff}} = \frac{\max(N_i^\mrm{eff})}{N_i^{\mrm{eff}}}~,
\end{equation}
which we refer to as ``linear'' reweighting. In this scheme, two
event samples at widely spaced energies, containing for instance 
10k and 1M events, respectively, would receive weights 100 and 1,
respectively. I.e., the power of the 
measurement with the \emph{lowest} statistics would be artificially
enhanced, in order for the global fit not to be totally dominated by
the higher-statistics one. This strategy should be considered 
extreme, however, since it grants no benefit at all to measurements
performed with superior statistics, and one therefore risks being overly
sensitive to noise in the poorly measured ones. As an intermediate 
compromise, we propose to let the samples enter with relative weights
\begin{equation}
w_i^{\mrm{eff}} = \sqrt{\frac{\max(N_i^\mrm{eff})}{N_i^{\mrm{eff}}}}~.
\end{equation}
such that the samples in the example above would enter with weights 
10 and 1, rather than 100 and 1. We refer to this as ``square root''
reweighting. 

Obviously, we do not intend this to define a rigorous procedure,
but view it as a first attempt at highlighting and addressing the disparate
statistical powers available in the various sets. 
\begin{table}[t]
    \centering
    \begin{tabular}{lcr|rr}\toprule
                 &                & & \multicolumn{2}{c}{ --------------------- \ 
        $N^\mrm{eff}_i$ \ ---------------------} \\ 
        Analysis & $\sqrt{s}$~[GeV] & $N_\text{evt}$ &$P(N_\mrm{ch})$ \&
        $\left<p_T\right>(N_\mrm{ch})$ &  $dN_\mrm{ch}/dp_\perp$  \\
        \midrule
        CDF 1988\cite{Abe:1988yu} & 630   &  9400  & & 47k\\
        CDF 2002\cite{Acosta:2001rm} & 630  & 1963157 &  2.0M\\
        UA5 1989\cite{Ansorge:1988kn} & 900   & 1189 &
        0.8M\\
        ATLAS 2010\cite{ATLAS:1266235} & 900   & 124782 & 0.8M & 162k\\
        CDF 1988\cite{Abe:1988yu} &   1800  & 55700 & & 9.2M \\
        CDF 2002\cite{Acosta:2001rm}  &   1800  & 2079558 &  11.2M\\
        CDF Run-\textsc{II}\cite{Aaltonen:2009ne,Run2:MBnote} &   1960  & 9788000 & 11.7M & 9.8M\\
        ATLAS 2010\cite{ATLAS:1266235} & 7000  & 5395000 & 5.4M & 5.4M\\
        \midrule
        $\max(N_i^\mrm{eff})$ & & & \multicolumn{2}{c}{11.7M} \\
        \bottomrule
    \end{tabular}\\[3mm]
{\footnotesize
\begin{tabular}{rr}
\multicolumn{1}{l}{\hspace*{0.75cm}$N_i^\mrm{eff}$ for $P(N_\mrm{ch})$
  \& $\left<p_T\right>(N_\mrm{ch})$} &
\multicolumn{1}{l}{\hspace*{0.75cm}$N_i^\mrm{eff}$ for $dN_\mrm{ch}/dp_\perp$} \\[-0.4cm] 
\includegraphics[scale=0.84]{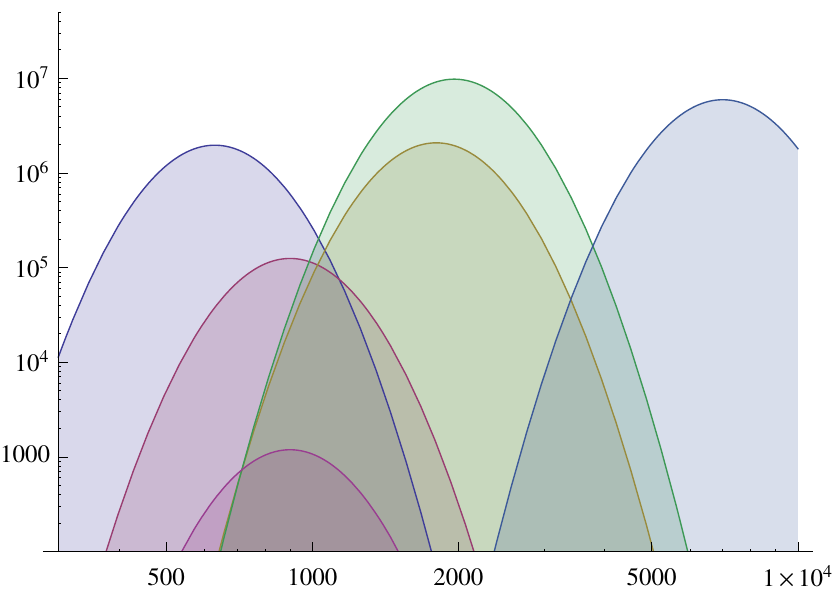}&
\includegraphics[scale=0.84]{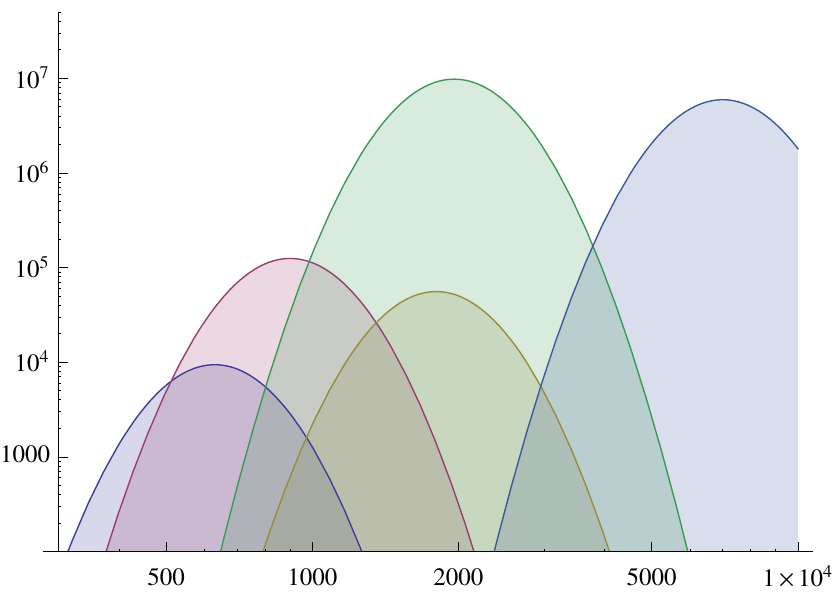}\\
$\sqrt{s}$ [GeV] & $\sqrt{s}$ [GeV]
\end{tabular}}
    \caption{{\sl Top Row:} Number of events contained in each data sample used, and
      the effective event numbers, $N_i^\mrm{eff}$, 
      for each observable, for each sample used for that
      observable. {\sl Bottom Row:} illustration of the
      relative sizes of each of the samples and their 
      contributions to  $N_i^\mrm{eff}$, according to
      eq.~(\ref{eq:Neff}), for {\sl Bottom Left:} $P(N_\mrm{ch})$ \&
      $\left<p_T\right>(N_\mrm{ch})$ and {\sl Bottom Right:} $dN_\mrm{ch}/dp_\perp$.}
    \label{tab:nevt}
\end{table}

As a final comment, we note that 
a certain freedom exists for the parameterization of the generator response
which allows for systematic checks of the validity of obtained tuning results
which can also be turned into typical spreads which we use as rough uncertainty
estimates in this study. These estimates and their properties are
described in greater 
detail in \cite{Buckley:2010nk}. We include them as light shaded
(cyan) bands on the plots in the following subsections, added in
quadrature with the ordinary $\chi^2$-based fit uncertainties computed
by Minuit, which are shown as darker shaded bands.

\subsection{Data Sets \label{sec:datasets}}

To study the energy dependence of parameters properly, we need to make
sure that the experimental data at each energy has either
been corrected for detector effects in a comprehensible way or that
the uncorrected data is presented with enough information on
efficiencies so that it can be 
used with \rivet. Further, the distributions we choose must be both sensitive to
the parameters we want to tune and, simultaneously, consistently available
at all collider energies with phase-space regions and trigger
conditions not too different from one to the next. 

A minimal set of minimum-bias distributions that matches these requirements
for our selection of tuning parameters and for energies ranging from
630 \GeV\ to 7\TeV\ is: 
\begin{itemize}
    \item The charged-particle multiplicity-distribution
      ($N_\text{ch}$) 
    \item The charged-particle \pt-distribution 
    \item The charged-particle $\langle \pt \rangle$
      vs. $N_\text{ch}$-distribution
\end{itemize}
We would have liked to extend the reach of our study by including data 
from the RHIC experiments but were unable to find any published
(corrected) data for the $\langle \pt \rangle$ vs. $N_\text{ch}$
distribution at 200 GeV. The same is true for older experiments such as
SFM, where multiplicities and \pt -distributions were measured in pp-reactions
at 62 \GeV and lower. As has recently been emphasized \cite{Wraight:2011ej},
 it would also have been interesting to add forward-backward
 correlations to the list of variables, since these are particularly
 sensitive to the mix of short- vs.\ long-distance processes, but
 in the energy range we consider, only UA5 has so far reported
 measurements \cite{Ansorge:1988fg}, with numbers at Tevatron and LHC
 energies not (yet)  available. 

\begin{table}[t!]
    \centering
\scalebox{1.0}{\small
\begin{tabular}{lrrlccc|crr}
\toprule
 & $\sqrt{s}$ &  &  & &
$p_{T\mrm{min}}$ & \multicolumn{2}{c}{Fitrange} &\multicolumn{2}{c}{Weight} \\ 
Exp/Trig & \hspace*{-3mm}[GeV] & Ref & Observable  & \hspace*{-1mm}$|\eta|_{\mrm{max}}$\hspace*{-1mm} &
[GeV] & $N_{\mrm{ch}}\ge 1$ & $N_{\mrm{ch}}\ge 6$ & \hspace*{-2mm}local & global \\ 
\midrule
% 630 GeV observables
\textbf{CDF}/TC1 & 630 & {\cite{Abe:1988yu}} &
 $dN/dp_\perp$ & $1.0$ & $0.4$ & $0.4$ -- 3 &  $0.4$ -- 3 &1.0 & 15.8\\
\textbf{CDF}/TC1 & 630 & \cite{Acosta:2001rm} &
 $P(N_{\mrm{ch}})$ & $1.0$ & $0.4$ & 1 -- 16 & 6 -- 16 & 1.0 & 2.4\\
 &&&$P(N_{\mrm{ch}})$ & $1.0$ & $0.4$ & 16 -- 30 & 16 -- 30 & 5.0 & 12.1\\
\textbf{CDF}/TC1 & 630 & \cite{Acosta:2001rm} &
 $\langle p_\perp \rangle$ vs.\ $N_{\mrm{ch}}$ & $ 1.0$ & $0.4$ 
 & 1 -- 16  & 6 -- 16 & 1.0 & 2.4\\
 &&&$\langle p_\perp \rangle$ vs.\ $N_{\mrm{ch}}$ & $ 1.0$ & $0.4$ 
 & 16 -- 30  & 16 -- 30 & 5.0 & 12.1\\
\midrule 
% 900 GeV observables --- UA5
\textbf{UA5}/TU1 & 900 & \cite{Ansorge:1988kn} & 
  $P(N_{\mrm{ch}})$& $0.5$ & $0.0$ &  1 -- 19 & 6 -- 19 &  1.0 & 3.8\\
\textbf{UA5}/TU1  & 900 & \cite{Ansorge:1988kn} & 
  $P(N_{\mrm{ch}})$ & $0.5$ & $0.0$ & 19 -- 30 & 19 -- 30 &  5.0 & 19.1\\
\midrule
% 900 GeV observables --- ATLAS
\textbf{ATL}/TA$^*$ & 900 & \cite{ATLAS:1266235} & 
  $dN/dp_\perp$ & $2.5$ & $0.5$ & 0.5 -- $10.0$ & 0.5 -- 10.0 & 1.0 & 8.5\\
\textbf{ATL}/TA$^*$ & 900 & \cite{ATLAS:1266235} & 
  $P(N_{\mrm{ch}})$ & $2.5$ & $0.5$ & 1
-- 25 & 6-- 25 &  1.0 & 3.8\\
\textbf{ATL}/TA$^*$ & 900 & \cite{ATLAS:1266235} & 
  $P(N_{\mrm{ch}})$& $2.5$ & $0.5$ & 25 -- 60 & 25 -- 60 & 5.0 & 19.1\\
\textbf{ATL}/TA$^*$ & 900 & \cite{ATLAS:1266235} & 
 $\langle p_\perp \rangle$ vs.\ $N_{\mrm{ch}}$  & $ 2.5$ & $0.5$ 
 & 1 -- 25  & 6 -- 25 & 1.0 & 3.8\\
 &&&$\langle p_\perp \rangle$ vs.\ $N_{\mrm{ch}}$  & $ 2.5$ & $0.5$ 
 & 25 -- 60  & 25 -- 60 & 5.0 & 19.1\\
\midrule 
% 1800 GeV observables
\textbf{CDF}/TC1 & 1800 & {\cite{Abe:1988yu}} &
 $dN/dp_\perp$ & $1.0$ & $0.4$ & $0.4$ -- $10.0$ &  $0.4$ -- $10.0$ & 1.0 & 1.1\\
\textbf{CDF}/TC1 & 1800 & \cite{Acosta:2001rm} &
 $P(N_{\mrm{ch}})$ & $1.0$ & $0.4$ & 1 -- 17  & 6 -- 17 & 1.0 & 1.0\\
\textbf{CDF}/TC1 & 1800 & \cite{Acosta:2001rm} &
 $P(N_{\mrm{ch}})$ & $1.0$ & $0.4$ & 17 -- 40 & 17 -- 40 & 5.0 & 5.0\\
\textbf{CDF}/TC1 & 1800 & \cite{Acosta:2001rm} &
 $\langle p_\perp \rangle$ vs.\ $N_{\mrm{ch}}$ & $ 1.0$ & $0.4$ 
 & 1 -- 17  & 6 -- 17 & 1.0 & 1.0\\
\textbf{CDF}/TC1 & 1800 & \cite{Acosta:2001rm} &
 $\langle p_\perp \rangle$ vs.\ $N_{\mrm{ch}}$ & $ 1.0$ & $0.4$ 
 & 17 -- 40  & 17 -- 40 & 5.0 & 5.0\\
\midrule 
% 1960 GeV observables
\textbf{CDF}/TC2 & 1960 & \cite{Aaltonen:2009ne} &
 $dN/dp_\perp$ & $1.0$ & $0.4$ & $0.4$ -- $15.0$ &  $0.4$ -- $15.0$ &1.0 & 1.0\\
\textbf{CDF}/TC2 & 1960 & \cite{Run2:MBnote} &
 $P(N_{\mrm{ch}})$ & $1.0$ & $0.4$ & 1 -- 18 & 6 -- 18 & 1.0 & 1.0 \\
\textbf{CDF}/TC2 & 1960 & \cite{Run2:MBnote} &
 $P(N_{\mrm{ch}})$ & $1.0$ & $0.4$ & 18 -- 30 & 18 -- 30 & 5.0 & 5.0\\
\textbf{CDF}/TC2 & 1960 & \cite{Aaltonen:2009ne} &
 $\langle p_\perp \rangle$ vs.\ $N_{\mrm{ch}}$ & $ 1.0$ & $0.4$ 
 & 1 -- 18  & 6 -- 18 & 1.0 & 1.0\\
\textbf{CDF}/TC2 & 1960 & \cite{Aaltonen:2009ne} &
 $\langle p_\perp \rangle$ vs.\ $N_{\mrm{ch}}$ & $ 1.0$ & $0.4$ 
 & 18 -- 30  & 18 -- 30 & 5.0 & 5.0 \\
\midrule
% ATLAS 7000 GeV observables
\textbf{ATL}/TA$^*$ & 7000 & \cite{ATLAS:1266235} & 
  $dN/dp_\perp$ & $2.5$ & $0.5$ & $0.5$ -- $40$ & $0.5$ -- $40.0$ & $1.0$ & 1.5 \\
\textbf{ATL}/TA$^*$ & 7000 & \cite{ATLAS:1266235} & 
  $P(N_{\mrm{ch}})$ & $2.5$ & $0.5$ & 1 -- 49 & 6 -- 49 &  1.0 & 1.5\\
\textbf{ATL}/TA$^*$ & 7000 & \cite{ATLAS:1266235} & 
  $P(N_{\mrm{ch}})$ & $2.5$ & $0.5$ & 49 -- 70 & $49$ -- $70$ & $5.0$ & 7.4 \\
\textbf{ATL}/TA$^*$ & 7000 & \cite{ATLAS:1266235} & 
 $\langle p_\perp \rangle$ vs.\ $N_{\mrm{ch}}$ & $ 2.5$ & $0.5$ 
 & 1 -- 49  & 6 -- 49 & 1.0 & 1.5\\
\textbf{ATL}/TA$^*$ & 7000 & \cite{ATLAS:1266235} & 
 $\langle p_\perp \rangle$ vs.\ $N_{\mrm{ch}}$ & $ 2.5$ & $0.5$ 
 & 49 -- 70  & 49 -- 70 & 5.0 & 7.4\\
\bottomrule
\end{tabular}}
    \caption{Observables and ranges included in the study. The trigger
      (``Trig'') conditions are as follows: {\sl TC1:} CDF Run
      I MB \cite{Abe:1988yu}, {\sl TC2:} CDF Run II MB\cite{Aaltonen:2009ne}, {\sl TU1}: UA5 MB \cite{Abe:1988yu}, 
      {\sl TA$^*$:} ATLAS trigger requiring
      $\ge 1$ ($\ge 6$) charged particles within $|\eta|<2.5$ and $p_\perp >
      0.5$~GeV for the $N_{\mrm{ch}}\ge 1$ ($N_{\mrm{ch}}\ge 6$) sample.  
    \label{tab:nch}. }
\end{table}
We consider two separate event samples: one where all events with at
least one charged track are included, corresponding to a conventional 
min-bias definition (though excluding the ``zero bin'' when comparing
to experiments that include it in their measurements), 
and one where only events with
$N_\mrm{ch}\ge 6$ are included, corresponding to a sample in which
diffractive contributions are expected to be strongly suppressed. Due
to the ambiguities discussed earlier concerning the treatment of 
diffraction, we 
shall base our conclusions mainly on the $N_\mrm{ch}\ge 6$ sample,
using the other as a further counter-check into the diffractive
region. Since measurements with an explicit $N_\mrm{ch}\ge 6$ definition
have so far only been carried out by ATLAS, the closest we can get 
for other experiments is to suppress the five first bins in
$P(N_{\mrm{ch}})$ and $\langle \pt \rangle$ vs. $N_\text{ch}$, and 
adjusting the normalization of the former such that the 
remaining bins sum to unity. 

A further complication concerns the phase-space (``fiducial'') 
regions measured by the different experiments. Although no two
experiments have exactly the same coverage, it is here of great
help that all of the experiments we consider have taken data at at
least two energies, and in the best cases also in several different phase
space regions. This effectively allows us to construct a kind of
bootstrapped path among the different regions. In fact, without such
counter-checks, the method we propose here would be badly compromised ---
 one could then never be certain whether a deviation in the optimized
parameters is caused by energy dependence or by the difference in 
phase space regions. We  therefore  encourage the RHIC,
Tevatron, and LHC experiments in their efforts to make measurements
using several different combinations of trigger conditions, phase
space regions, and collider energies. Ultimately, it is by such 
comprehensive and systematic sets of measurements, and by the 
comprehensive tests that they enable, that we may establish a truly
reliable modeling of collider final states. 

The observables, ranges, and weights used for both the $N_{\mrm{ch}}\ge 1$
and $N_{\mrm{ch}}\ge 6$ samples are given in table
\ref{tab:nch}. Larger weights are given to the high-multiplicity tails
of the $P(N_\mrm{ch})$ and $\left<p_\perp\right>(N_\mrm{ch})$ 
distributions to emphasize their asymptotic slopes. As our
definition for ``high multiplicity'', we took the $N_\mrm{ch}$ value
that came closest to separating out the 1\% highest-multiplicity
events, for each measurement.

\section{Consistency of Energy Scaling \label{sec:energyscaling}}

In this section, we study the degree to which the parameters obtained
for a best-fit tune across all included data sets and collider
energies are consistent with those obtained when we include only
specific subsets of the data. In particular, we focus on the
consistency on the assumed energy scaling by comparing the results of
the global fit to results obtained at each energy separately. 

We study three specific questions
\begin{itemize}
\item Is the assumed scaling law governing the infrared regularization
  scale for multiple parton interactions consistent with what one
  finds when optimizing the tuning at each energy separately?
\item Is the transverse mass distribution of the proton
  (assumed unchanging with energy) consistent with what one finds when
  optimizing the tuning at each energy separately? 
\item Is the assumed color reconnection strength 
  (assumed unchanging with energy) consistent with what one finds when
  optimizing the tuning at each energy separately? 
\end{itemize}

\begin{figure}[tp]
\subfigure[\label{fig:p82nch1} \parp{82} vs \sqrts, $N_\text{ch}\geq 1$ ]{
\includegraphics[width=0.5\textwidth]{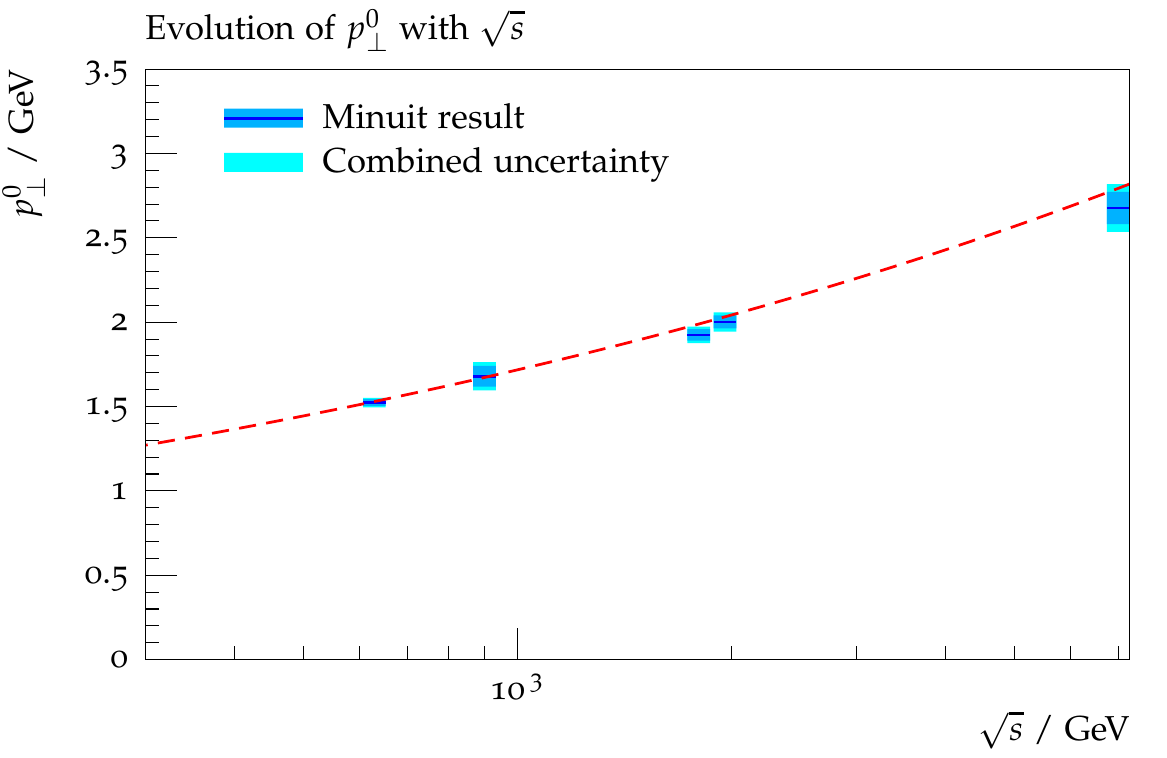}
}
\subfigure[\label{fig:p82nch6} \parp{82} vs \sqrts, $N_\text{ch}\geq 6$ ]{
\includegraphics[width=0.5\textwidth]{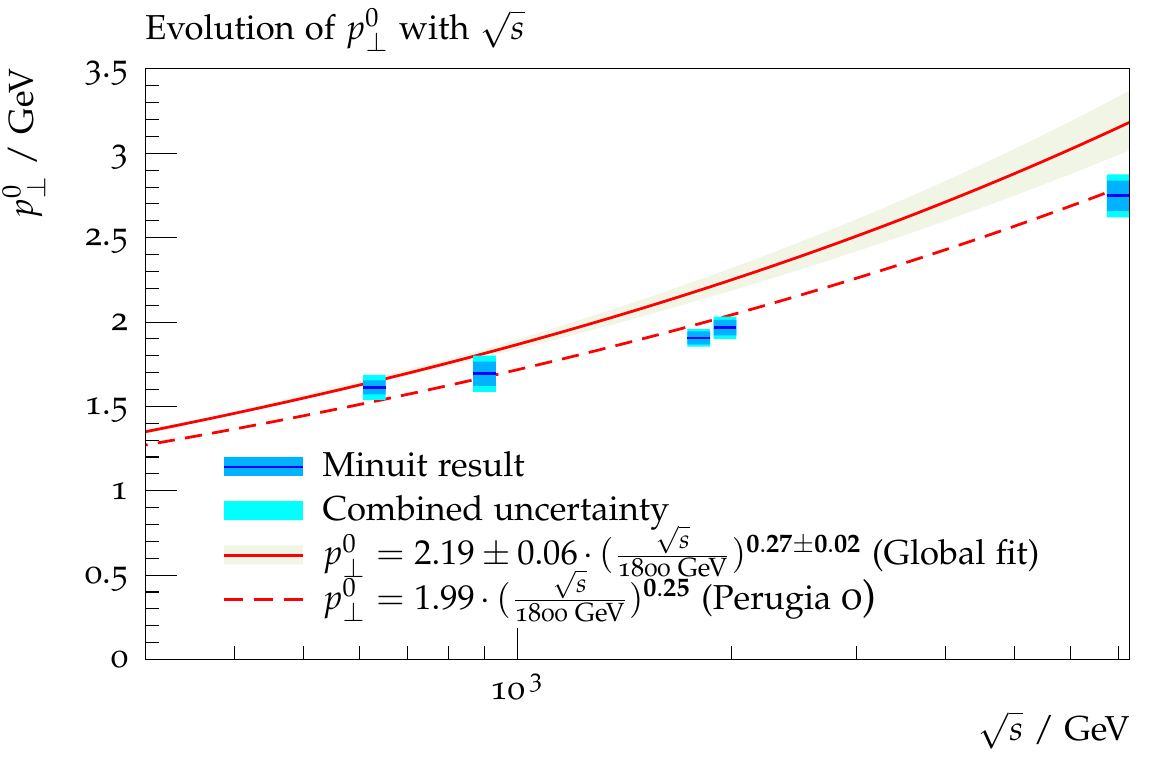}
}\\
\subfigure[\label{fig:p83nch1} \parp{83} vs \sqrts, $N_\text{ch}\geq 1$ ]{
\includegraphics[width=0.5\textwidth]{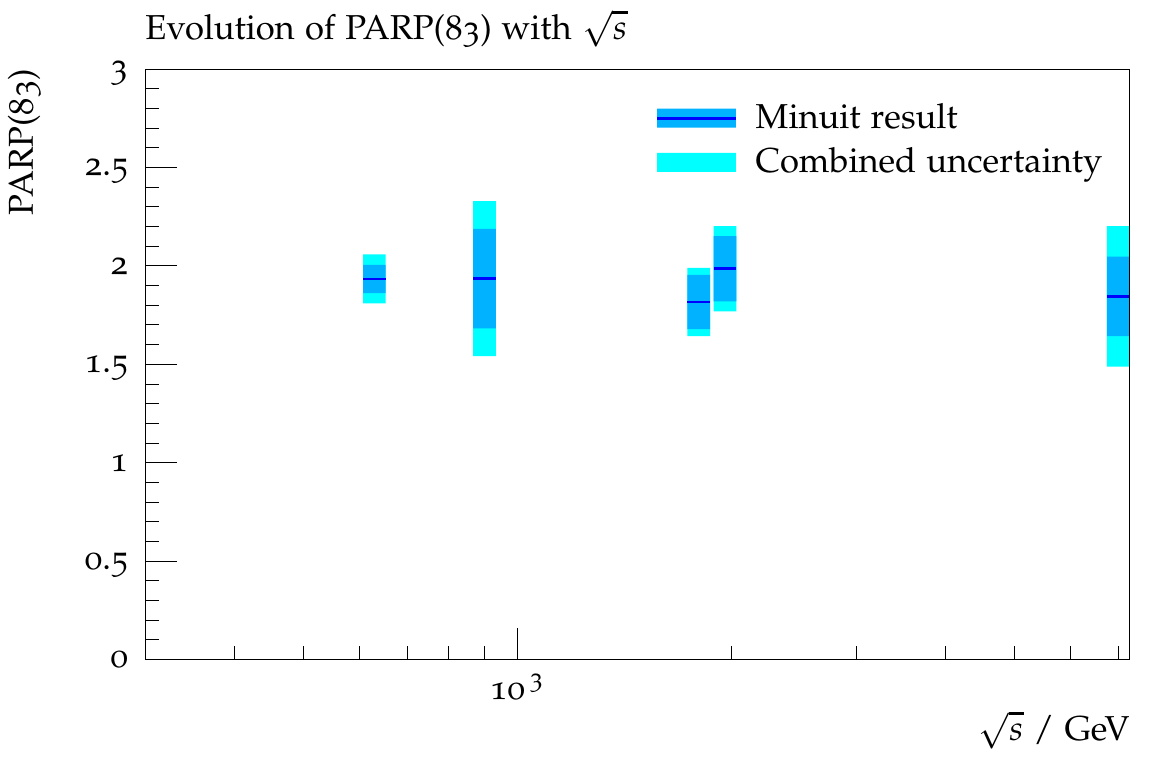}
}
\subfigure[\label{fig:p83nch6} \parp{83} vs \sqrts, $N_\text{ch}\geq 6$ ]{
\includegraphics[width=0.5\textwidth]{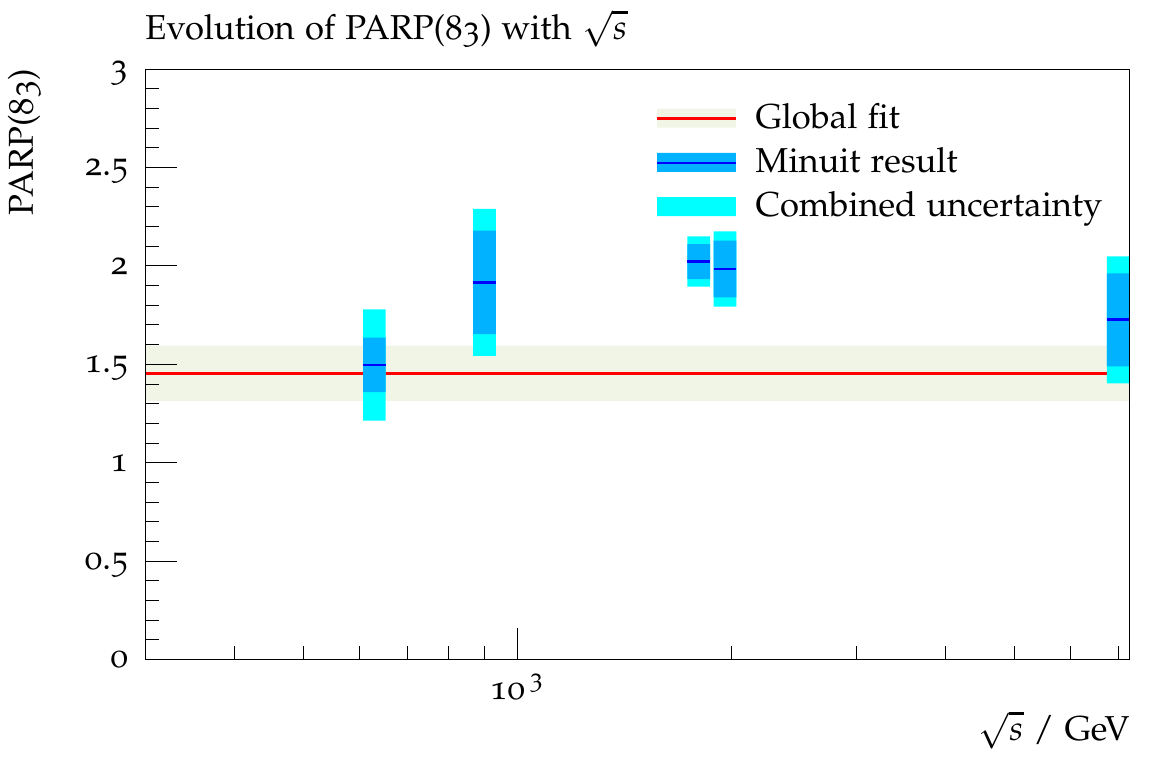}
}
\subfigure[\label{fig:p78nch1} \parp{78} vs \sqrts, $N_\text{ch}\geq 1$ ]{
\includegraphics[width=0.5\textwidth]{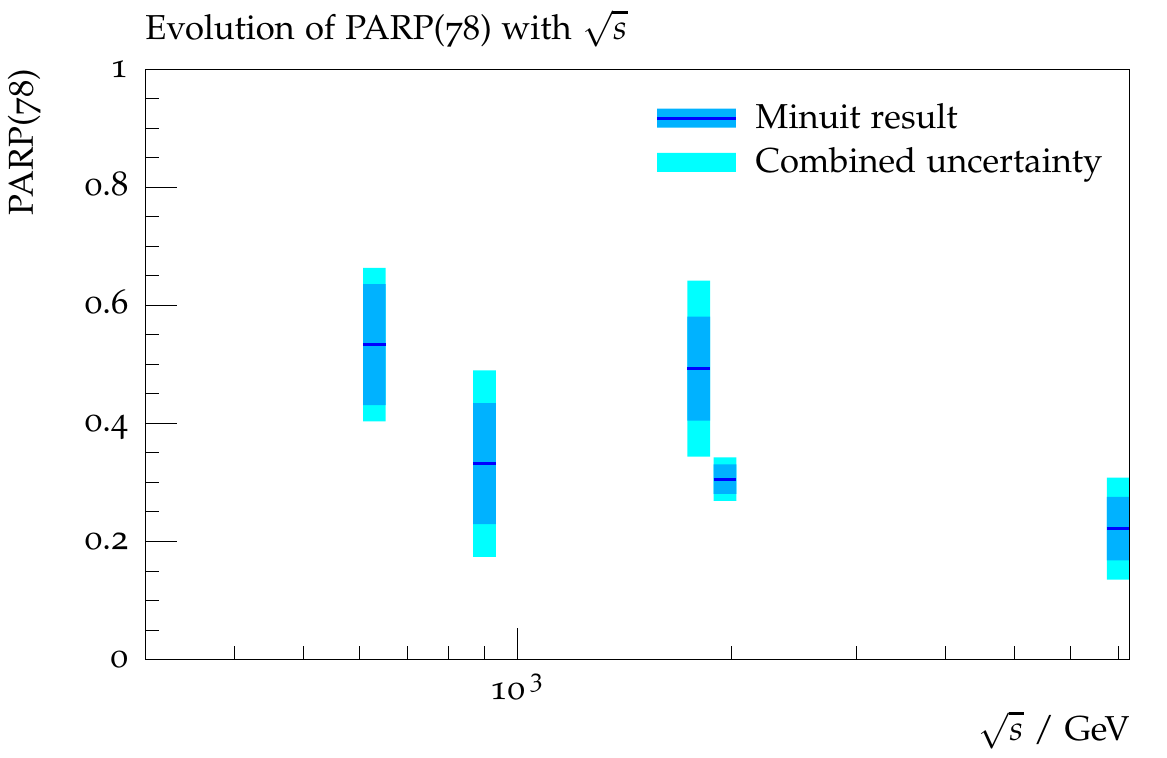}
}
\subfigure[\label{fig:p78nch6} \parp{78} vs \sqrts, $N_\text{ch}\geq 6$ ]{
\includegraphics[width=0.5\textwidth]{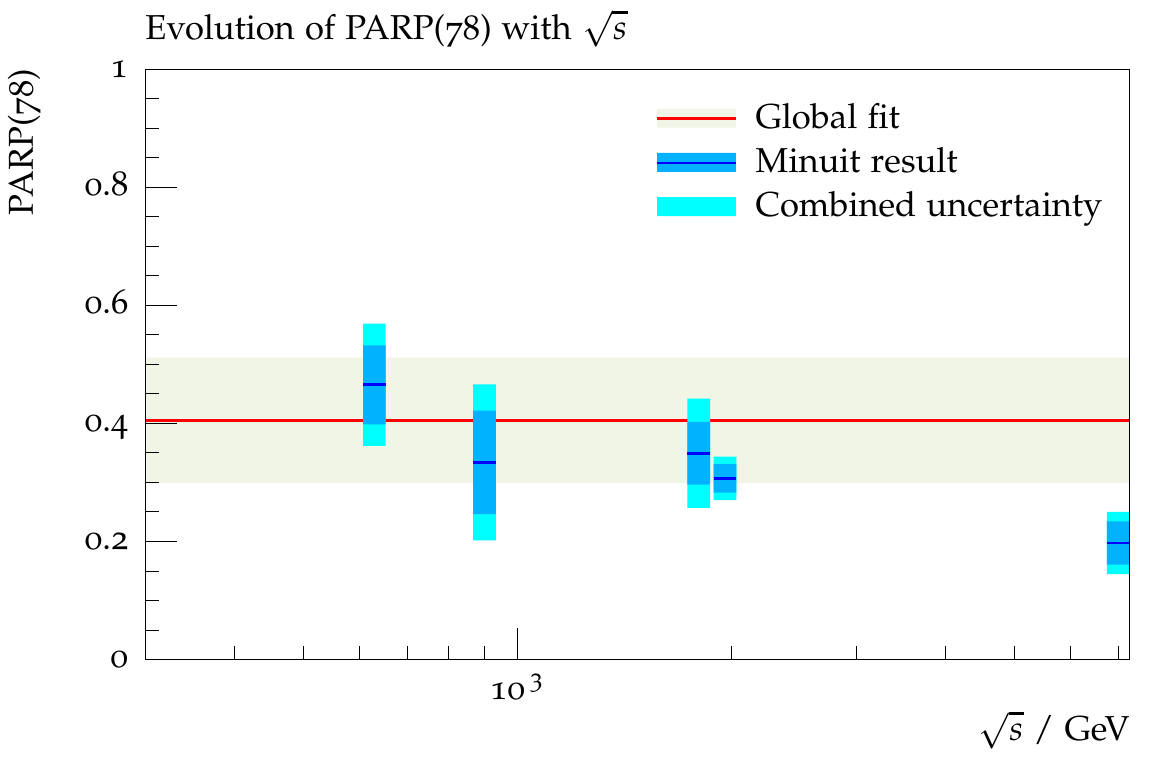}
}
\caption{Energy dependence of the the three tune parameters, from top
  to bottom: \parp{82}, \parp{83}, and \parp{78}. 
  Independent optimizations (blue/shaded lines)  
compared to global fit curve (red solid curves). {\sl Left:}
$N_{\mrm{ch}}\ge 1$ sample. {\sl Right:} $N_{\mrm{ch}}\ge 6$ sample. 
\label{fig:parp82}}
\end{figure}

The evolution of the infrared regularization scale with energy is
depicted in \FigRef{fig:parp82}, with the left-hand panes showing the 
results for the $N_{\mrm{ch}}\ge1$ sample 
and the right-hand panes those for $N_{\mrm{ch}}\ge 6$. The scaling of
Perugia 0 (red dashed lines) is compared to the 
global fit (red solid line in light shaded band) and to the 
independent optimizations (blue horizontal lines inside cyan bands). 
As mentioned in \SecRef{sec:professor}, the inner (darker blue) bands
correspond to the fit parameter uncertainties calculated by Minuit and
the outer (lighter cyan) bands include an estimate of 
\textsc{Professor}'s interpolation uncertainty \cite{Buckley:2010nk} 
as well, with the two
added in quadrature.

We make four conclusions concerning \parp{82}. One, 
that the results of the independent optimizations are consistent with
the functional form represented by \EqRef{eqn:ptmin} 
and hence we find no evidence for a need for any
significant departure from the model assumptions in this
energy range.  Two, that the different individual
data sets appear to be consistent and reconcilable within this modeling
context --- the two different CDF measurements, the left- and
right-hand sample definitions, and the ATLAS and UA5 measurements at
900 GeV, all appear to give consistent parameters. 
Three, that the light shaded (cyan) bands are very small, and that the
tune result can therefore be considered technically stable. And
finally, that the global fit \emph{does not} coincide with the
independent optimizations. Although not huge, this deviation hints
that one or more of the other parameters must be exhibiting a
non-universal behaviour. 

Turning now to the scaling of the transverse shape parameter, \parp{83},
this is particularly interesting since it
could reveal whether minimum-bias collisions at different energies
effectively probe a different ``average proton shape''. Such a
variation could, e.g., be generated by correlations between $b$ and $x$
 (see, e.g., \cite{Pancheri:2010dg,Blok:2010ge,Frankfurt:2010ea,Corke:2011yy}), 
folded with the different $x$ ranges that are accessible at each
energy. Roughly speaking, we might then expect to see a slightly 
more lumpy average proton at lower energies, consistent with a higher average
$x$ at those energies, and a smoother proton at higher energies / lower
average $x$. Results of the local and global tunes for \parp{83} 
are shown in the middle panes of \FigRef{fig:parp82}.

In our main $N_\mrm{ch}\ge 6$ sample, shown in the right-hand pane of
\FigRef{fig:parp82}, there may be  some weak
evidence for such a trend, with a close-to-Gaussian proton (\parp{83}=2) favoured
by the high-energy data and a slightly more peaked distribution
favoured by the 630-GeV data. However, note that the shaded bands are here
larger, indicating that the fit is less well constrained than
it was for \parp{82}. Also, when we include the lowest-multiplicity
bins, in the left-hand pane, the trend disappears. 
Our tentative conclusion is therefore that the uncertainties are
too large to make any firm conclusions, but that the model at least 
appears to be self-consistent within those uncertainties. 
Further studies at lower energies (e.g., including $pp$
data from RHIC and/or further Tevatron studies at 630 GeV) 
and/or attempting to isolate different effective $x$ ranges at higher
energies, e.g., by using different rapidity and/or trigger regions, could
contribute significantly to probing this question further. 
More theoretical work to improve the understanding of the relationship
between the language used here and that of other phenomenological
models, as was done, e.g., by \cite{Rogers:2009ke}, would also be
valuable and could open the possibility for a 
consistent ``importation'' of constraints
from related physical models into the Monte Carlo context.

Returning to the present study, note also that \professor furnishes us with one
additional key piece of information. When optimizing several
parameters simultaneously, we not only get the optimized values for
each parameter separately; we also get a correlation matrix between
them. When interpreting our results, one should therefore be aware 
that there is a  strong correlation between \parp{82} and \parp{83}. 
Hence, there is still a possibility that \parp{83} could
have a more significant energy dependence, to be traded off against that of
\parp{82}. However, since the fit result for \parp{82} was, itself,
quite stable, we consider this possibility something of a minority
report, not favored by the central fits; but also 
not completely excluded.

Finally, the Color Reconnection strength, \parp{78}, is --- perhaps
not surprisingly --- the least well constrained parameter, with the
individual fit results showing a preferred scaling with energy that is not
accounted for by the underlying model. Given the large uncertainties
surrounding color correlations and final-state interactions in hadron
collisions, this can be considered a reminder that, although the
models do make attempts at incorporating this kind of phenomena, our
understanding is still very far from complete or reliable. For
the time being, any global tuning relying on the CR models considered
in this study would be forced to make a compromise
between the high- and low-energy data. Pragmatic alternatives for
physics studies at specific colliders would
range from giving that particular energy a larger weight in the global
fit to simply abandoning a global fit altogether. Although clearly not
theoretically satisfactory, the latter may be a useful strategy for
applications in which the Monte Carlo modeling is only used as a
sophisticated differential ``parameterization'' of the behavior of the
data. In particular at 7 TeV, large data samples  are now
becoming available that probe many different and
complementary phase space regions in detail, allowing a fairly
complete set of constraints to be obtained for that particular
collider energy. 

\begin{table}[t]
\centering
\begin{tabular}{l*{6}{c}}
\toprule
\sqrts & 630 \GeV & 900 \GeV & 1.8 \TeV & 1.96 \TeV & 7 \TeV & Global fit\\
\parp{89} & 630.0 & 900.0 & 1800.0 & 1960.0 & 7000.0 & 1800.0\\
\midrule
\multicolumn{6}{l}{Tuning to observables with $N_\text{ch}\geq1$}\\
\parp{78} & $0.53\pm0.10$  & $0.33\pm0.10$ & $0.49\pm0.09$ & $0.31\pm0.03$ & $0.22\pm0.05$\\
\parp{82} & $1.52\pm0.02$  & $1.68\pm0.06$ & $1.92\pm0.03$ & $2.00\pm0.04$ & $2.68\pm0.10$\\
\parp{83} & $1.93\pm0.07$  & $1.94\pm0.25$ & $1.82\pm0.14$ & $1.99\pm0.17$ & $1.85\pm0.20$\\
\midrule
\multicolumn{6}{l}{Tuning to observables with $N_\text{ch}\geq6$}\\
\parp{78} &  $0.47\pm0.07$ & $0.33\pm0.09$ & $0.35\pm0.05$ & $0.31\pm0.02$ & $0.20\pm0.04$ & $0.40\pm0.11$\\
\parp{82} &  $1.61\pm0.04$ & $1.69\pm0.07$ & $1.91\pm0.04$ & $1.97\pm0.05$ & $2.75\pm0.01$ & $2.19\pm0.06$ \\
\parp{83} &  $1.50\pm0.14$ & $1.92\pm0.03$ & $2.02\pm0.09$ & $1.98\pm0.01$ & $1.73\pm0.02$ & $1.45\pm0.14$\\
\parp{90} &                &               &               &               &               & $0.27\pm 0.02$ \\
\bottomrule
\end{tabular}
%\caption{Tuning results for individual energies. Shown are the mean and standard deviation of results obtained from 70 parameterisations.}
\caption{Tuning results obtained with the maximum information
  parameterisations. The errors quoted are those calculated by Minuit.}
\label{tab:tunedparams}
\end{table}

\section{Conclusions \label{sec:conc}}

We have argued that the capabilities of modern tuning tools
 can and should be used for more than just making ``best
fits'' to a collection of data. For example, 
by making independent optimizations of the
MC generator parameters for several different collider energies, we
have here obtained a data-driven test of the universality of the generator
modeling. Three of the most important generator parameters controlling
the underlying-event and minimum-bias physics in the \textsc{Pythia 6}
generator were included in the study, corresponding to: the infrared
regularization scale for multiple parton interactions (MPI), the proton
transverse shape, and the strength of color reconnections (CR). A
brief discussion of each of these parameters was given in
\SecRef{sec:model}. The \professor tool used for the tunings as well
as a weighting strategy that attempts to take the size of 
disparate statistical samples and measurements at closely spaced
energies formally into account were described in
\SecRef{sec:professor}. The data sets consisted of minimum-bias 
measurements of charged
particle multiplicities, $p_\perp$ spectra, and the average of 
$p_{\perp}$ vs.\ multiplicity, as described in \SecRef{sec:datasets}.

Our numerical results were presented and discussed 
in \SecRef{sec:energyscaling}. We find
that the result of independent optimizations of the IR regularization
scale, energy by energy, are consistent with the power-law behaviour assumed
by the model, at least within the energy range we were able to probe,
from 630 to 7000 GeV. The transverse matter distribution may exhibit mild
deviations from universality, a question which data in particular from
minimum-bias measurements at RHIC could help shed further light
on. Finally, the optimal value of the color-reconnection strength
appears to vary significantly with energy, with lower values preferred
at higher energies, in contrast to the intrinsic assumption of a constant
strength in the model. This confirms the theoretical evaluation, that 
the CR modeling is currently the largest source of theoretical
ambiguity, and
emphasizes that at least the models investigated here cannot be
considered truly universal over the studied energy range. 
  
The parameter values corresponding to each of our ``local'' (i.e.,
energy-by-energy) tunes as well as those of a ``global'' one are collected
in \TabRef{tab:tunedparams}. All other parameter values were taken to
be those of the Perugia 0 tune \cite{Skands:2010ak}. These new tunes
have been included in \textsc{Pythia} starting from version 6.4.25,
with tune numbers 360 --- 365 (see the \textsc{Pythia} update notes
for details).

Finally, we argue that procedures similar to the one followed here 
can be used to advantage also in other contexts, to give a clearer picture
of which regions the modeling is able to describe with approximately
universal parameters, which in turn helps isolate the 
genuinely problematic areas more easily. The trend of the optimized parameters to deviate in one or another
direction in the problematic regions may also give clues as to the root of
the problem, though such interpretations should be made in conjunction
with a good understanding of the correlations between the parameters
and a careful evaluation of the possible missing physics components in the
modeling. 
One clear possibility to apply this type of strategy to present
measurements would be
to perform independent optimizations using different complementary phase
space regions at each energy.

\section*{Acknowledgments}

This work was supported in part by the Marie Curie research training
network ``MCnet'' (contract number MRTN-CT-2006-035606). 
HS acknowledges support from the German Research Foundation (DFG) in
the framework of Graduate School 1504. 

\bibliographystyle{h-physrev3}
\bibliography{energy-scaling,data}

\begin{thebibliography}{10}

\bibitem{Buckley:2009bj}
A.~Buckley, H.~Hoeth, H.~Lacker, H.~Schulz, and J.~E. von Seggern,
\newblock Eur.Phys.J. {\bf C65}, 331 (2010), 0907.2973.

\bibitem{Sjostrand:2004pf}
T.~Sj{\"o}strand and P.~Z. Skands,
\newblock JHEP {\bf 03}, 053 (2004), hep-ph/0402078.
%%CITATION = HEP-PH/0402078;%%

\bibitem{Sjostrand:2004ef}
T.~Sj{\"o}strand and P.~Z. Skands,
\newblock Eur. Phys. J. {\bf C39}, 129 (2005), hep-ph/0408302.
%%CITATION = HEP-PH/0408302;%%

\bibitem{Sjostrand:2006za}
T.~Sj{\"o}strand, S.~Mrenna, and P.~Z. Skands,
\newblock JHEP {\bf 05}, 026 (2006), hep-ph/0603175.
%%CITATION = HEP-PH/0603175;%%

\bibitem{Bartalini:2010su}
P.~Bartalini, (ed.~) {\em et~al.},
\newblock (2010), 1003.4220.
%%CITATION = 1003.4220;%%

\bibitem{Skands:2010ak}
P.~Z. Skands,
\newblock Phys.Rev. {\bf D82}, 074018 (2010), 1005.3457.

\bibitem{Sjostrand:2007gs}
T.~Sj{\"o}strand, S.~Mrenna, and P.~Z. Skands,
\newblock Comput. Phys. Commun. {\bf 178}, 852 (2008), 0710.3820.
%%CITATION = 0710.3820;%%

\bibitem{Sjostrand:1987su}
T.~Sj{\"o}strand and M.~van Zijl,
\newblock Phys.Rev. {\bf D36}, 2019 (1987).

\bibitem{Ryskin:2011qe}
M.~Ryskin, A.~Martin, and V.~Khoze,
\newblock (2011), 1102.2844.

\bibitem{Buckley:2011ms}
A.~Buckley {\em et~al.},
\newblock (2011), 1101.2599.

\bibitem{Lai:1999wy}
CTEQ~Collaboration, H.~L. Lai {\em et~al.},
\newblock Eur. Phys. J. {\bf C12}, 375 (2000), hep-ph/9903282.
%%CITATION = HEP-PH/9903282;%%

\bibitem{Pumplin:2002vw}
CTEQ~Collaboration, J.~Pumplin {\em et~al.},
\newblock JHEP {\bf 07}, 012 (2002), hep-ph/0201195.
%%CITATION = HEP-PH/0201195;%%

\bibitem{Sherstnev:2007nd}
A.~Sherstnev and R.~S. Thorne,
\newblock Eur. Phys. J. {\bf C55}, 553 (2008), 0711.2473.
%%CITATION = 0711.2473;%%

\bibitem{Donnachie:1992ny}
A.~Donnachie and P.~V. Landshoff,
\newblock Phys. Lett. {\bf B296}, 227 (1992), hep-ph/9209205.
%%CITATION = HEP-PH/9209205;%%

\bibitem{Sandhoff:2005jh}
M.~Sandhoff and P.~Z. Skands,
\newblock (2006),
\newblock in hep-ph/0604120.

\bibitem{Buttar:2006zd}
C.~Buttar {\em et~al.},
\newblock (2006), hep-ph/0604120.
%%CITATION = HEP-PH/0604120;%%

\bibitem{Skands:2007zg}
P.~Z. Skands and D.~Wicke,
\newblock Eur. Phys. J. {\bf C52}, 133 (2007), hep-ph/0703081.
%%CITATION = HEP-PH/0703081;%%

\bibitem{Rathsman:1998tp}
J.~Rathsman,
\newblock Phys. Lett. {\bf B452}, 364 (1999), hep-ph/9812423.
%%CITATION = HEP-PH/9812423;%%

\bibitem{Andersson:1983jt}
B.~Andersson, G.~Gustafson, and B.~S{\"o}derberg,
\newblock Z. Phys. {\bf C20}, 317 (1983).
%%CITATION = ZEPYA,C20,317;%%

\bibitem{Andersson:1985qr}
B.~Andersson, G.~Gustafson, and B.~S{\"o}derberg,
\newblock Nucl. Phys. {\bf B264}, 29 (1986).
%%CITATION = NUPHA,B264,29;%%

\bibitem{Aaltonen:2009ne}
CDF~Collaboration, T.~Aaltonen {\em et~al.},
\newblock Phys. Rev. {\bf D79}, 112005 (2009), 0904.1098.
%%CITATION = 0904.1098;%%

\bibitem{Aad:2010rd}
ATLAS~Collaboration, G.~Aad {\em et~al.},
\newblock Phys. Lett. {\bf B688}, 21 (2010), 1003.3124.
%%CITATION = 1003.3124;%%

\bibitem{Waugh:2006ip}
B.~M. Waugh {\em et~al.},
\newblock (2006), hep-ph/0605034.
%%CITATION = HEP-PH/0605034;%%

\bibitem{rivet}
A.~Buckley {\em et~al.},
\newblock (2010), 1003.0694.
%%CITATION = 1003.0694;%%

\bibitem{Dobbs:2001ck}
M.~Dobbs and J.~B. Hansen,
\newblock Comput.Phys.Commun. {\bf 134}, 41 (2001).

\bibitem{Buckley:2010jn}
A.~Buckley and M.~Whalley,
\newblock (2010), 1006.0517.

\bibitem{Buckley:2010nk}
A.~Buckley {\em et~al.},
\newblock (2010), 1005.5357.

\bibitem{Wraight:2011ej}
K.~Wraight and P.~Skands,
\newblock (2011), 1101.5215,
\newblock Submitted to EPJC.

\bibitem{Ansorge:1988fg}
UA5~Collaboration, R.~Ansorge {\em et~al.},
\newblock Z.Phys. {\bf C37}, 191 (1988).

\bibitem{Abe:1988yu}
CDF~Collaboration, F.~Abe {\em et~al.},
\newblock Phys. Rev. Lett. {\bf 61}, 1819 (1988).
%%CITATION = PRLTA,61,1819;%%

\bibitem{Acosta:2001rm}
CDF~Collaboration, D.~E. Acosta {\em et~al.},
\newblock Phys. Rev. {\bf D65}, 072005 (2002).
%%CITATION = PHRVA,D65,072005;%%

\bibitem{Ansorge:1988kn}
UA5~Collaboration, R.~E. Ansorge {\em et~al.},
\newblock Z. Phys. {\bf C43}, 357 (1989).
%%CITATION = ZEPYA,C43,357;%%

\bibitem{ATLAS:1266235}
{ATLAS Collaboration},
\newblock CERN Report No. ATLAS-COM-CONF-2010-031, 2010 (unpublished),
\newblock (Was originally 'ATL-COM-PHYS-2010-268').

\bibitem{Run2:MBnote}
CDF~Collaboration, N.~Moggi {\em et~al.},
\newblock PUB-NOTE {\bf 9936} (2009).

\bibitem{Pancheri:2010dg}
G.~Pancheri, A.~Grau, R.~M. Godbole, and Y.~N. Srivastava,
\newblock (2010), 1007.0208.

\bibitem{Blok:2010ge}
B.~Blok, Y.~Dokshitzer, L.~Frankfurt, and M.~Strikman,
\newblock (2010), 1009.2714.

\bibitem{Frankfurt:2010ea}
L.~Frankfurt, M.~Strikman, and C.~Weiss,
\newblock (2010), 1009.2559.

\bibitem{Corke:2011yy}
R.~Corke and T.~Sj{\"o}strand,
\newblock (2011), 1101.5953.

\bibitem{Rogers:2009ke}
T.~Rogers and M.~Strikman,
\newblock Phys.Rev. {\bf D81}, 016013 (2010), 0908.0251.

\end{thebibliography}

\end{document}